\documentclass[%
twocolumn,
groupedaddress,
superscriptaddress,
amsmath,amssymb,
aps,
prx,
]{revtex4-1}

\usepackage{graphicx}
\usepackage{amsfonts}
\usepackage{bm}
\usepackage{hyperref,url}
\usepackage{changes}
\usepackage{xcolor}
\usepackage{soul}

\newcommand{\bd}{\begin{displaymath}}
\newcommand{\ed}{\end{displaymath}}
\newcommand{\be}{\begin{equation}}
\newcommand{\ee}{\end{equation}}
\newcommand{\bs}{\begin{subequations}}
\newcommand{\es}{\end{subequations}}
\newcommand{\ba}{\begin{eqnarray}}
\newcommand{\ea}{\end{eqnarray}}

\begin{document}

\title{Quantum Zermelo problem for general energy resource bounds}

\author{Josep Maria Bofill}
\email{jmbofill@ub.edu}
\affiliation{Departament de Qu\'{i}mica Inorg\`anica i Org\`anica,
Secci\'o de Qu\'{i}mica Org\`anica,
Universitat de Barcelona, Mart\'{i} i Franqu\`es 1, 08028 Barcelona, Spain}
\affiliation{Institut de Qu\'{i}mica Te\`orica i Computacional (IQTCUB),
Universitat de Barcelona, Mart\'{i} i Franqu\`es 1, 08028 Barcelona, Spain}

\author{\'Angel S. Sanz}
\email{a.s.sanz@fis.ucm.es}
\affiliation{Department of Optics, Faculty of Physical Sciences,
Universidad Complutense de Madrid,
Pza.\ Ciencias 1, Ciudad Universitaria -- 28040 Madrid, Spain}

\author{Guillermo Albareda}
\email{guillealpi@gmail.com}
\affiliation{Max Planck Institute for the Structure and Dynamics of Matter and Center for
Free-Electron Laser Science, Luruper Chaussee 149, 22761 Hamburg, Germany}
\affiliation{Institut de Qu\'{i}mica Te\`orica i Computacional (IQTCUB),
Universitat de Barcelona, Mart\'{i} i Franqu\`es 1, 08028 Barcelona, Spain}

\author{Ib\'{e}rio de P.R. Moreira}
\email{i.moreira@ub.edu}
\affiliation{Departament de Ci\`encia de Materials i Qu\'{i}mica F\'{i}sica,
Secci\'o de Qu\'{i}mica Org\`anica,
Universitat de Barcelona, Mart\'{i} i Franqu\`es 1, 08028 Barcelona, Spain}
\affiliation{Institut de Qu\'{i}mica Te\`orica i Computacional (IQTCUB),
Universitat de Barcelona, Mart\'{i} i Franqu\`es 1, 08028 Barcelona, Spain}

\author{Wolfgang Quapp}
\email{quapp@math.uni-leipzig.de}
\affiliation{Mathematisches Institut, Universit\"{a}t Leipzig, PF 100920, D-04009 Leipzig, Germany}

\date{\today}

\begin{abstract}
A solution to the quantum Zermelo problem for control Hamiltonians with general energy resource bounds is provided.
Interestingly, the energy resource of the control Hamiltonian and the control time define a pair of conjugate variables that minimize the energy-time uncertainty relation.
The resulting control protocol is applied to a single qubit as well as to a two-interacting qubit system represented by a Heisenberg spin dimer.
For these low-dimensional systems, it is found that physically realizable control Hamiltonians exist only for certain, quantized, energy resources. 
\end{abstract}

\maketitle

%%%%%%%%%%%%%%%%%%%%%%%%%%%%%%%%%%%%%%%%%%%%%%%%%%%%%%%%%%%%%%%%%%%%%%%
%%%%%%%%%%%%%%%%%%%%%%%%%%%%%%%%%%%%%%%%%%%%%%%%%%%%%%%%%%%%%%%%%%%%%%%

\section{Introduction}
\label{sec1}

Nature requires a quantum description \cite{horo4x09}.
Nonetheless, quantum-classical correspondence arguments are still in fashion because of their
usefulness to understand and explain the behavior of quantum systems \cite{olivar-bk}, and
also to devise new strategies to tackle quantum problems, as in the case of optimal control
strategies \cite{tannor:FaradayTrans:1986,gordon:ARPC:1997,brumer:AAMOP:2000,dalessandro-bk,dong:controlth:2010}.
In general, control scenarios are related either with the way of constructing a so-called
control Hamiltonian or with the procedure aimed at getting an appropriate initial ansatz
that, with time, evolves into the desired final quantum state.

From a mathematical point of view, any problem in quantum mechanics essentially consists in a more or less complete and precise construction of a unitary operator $\hat{U}(t,t_0)$, parametrically dependent on $t$, to represent the time evolution of the system between its energetically available mechanical states at $t_0$ and $t$. In the last years much attention has been devoted to the problem of finding
optimal unitary operators, $\hat{U}(t_f,t_i)$, which lead a given initial state $|\psi_i\rangle$,
at $t_i$, to another different but previously fixed final state $|\psi_f\rangle$, at $t_f$,
in the shortest possible time, $\Delta T = t_f - t_i$, under some constraining conditions.
Since finding the optimal unitary operator is equivalent to finding an optimal Hamiltonian,
two different routes can be explored.
On the one hand, if the constraint implies a bound in the energy resource, then the optimal
Hamiltonian is going to be time-independent and can easily be constructed by noting that the
corresponding unitary transformation $\hat{U}$ describes the shortest time evolution.
On the other hand, the constraint might imply a search for a time-dependent Hamiltonian
minimizing the time evolution, which means that it has to be determined and characterized by
variational approaches \cite{xwang15}.

The latter case is particularly relevant to those situations where the evolution of the
quantum system is either predetermined or inherently affected by an external field out
of our control (for instance, in problems within the scope of the quantum technologies).
Yet it would be desirable to take the system from one state to another one that
does not correspond to the natural evolution of such a system.
That is, if $\hat{U}_0$ describes such a natural evolution, it is of much interest to devise
a method or protocol that warrants the evolution from $|\psi_i\rangle$ to $|\psi_f\rangle$ in
the least time, provided that $|\psi_f\rangle \ne \hat{U}_0 |\psi_i\rangle$.
By invoking the aforementioned quantum-classical correspondence, this is actually the
quantum analog of the well-known classical Zermelo navigation problem \cite{zermelo31,caratheodory}.

Brody and Meier \cite{brody15} have investigated this problem in the field of quantum
processing.
More specifically, assuming that the quantum system is described by a bare background Hamiltonian,
$\hat{H}_0$, these authors established a method to obtain a time-optimal control Hamiltonian,
$\hat{H}_c(t)$, such that its combined action with $\hat{H}_0$, i.e.,
\be
 \hat{H}(t) = \hat{H}_0 + \hat{H}_c(t) ,
 \label{hamilton0}
\ee
generates a time-optimal unitary evolution from $|\psi_i\rangle$ to $|\psi_f\rangle$.
The protocol devised by these authors thus includes three key elements:
\begin{itemize}
 \item[(i)] A time-independent, bare background Hamiltonian, $\hat{H}_0$, which describes
 the natural evolution of the quantum system.

 \item[(ii)] A time-dependent control $\hat{H}_c(t)$ satisfying at any time the energy
 resource bound
 \be
  {\rm tr}\!\left(\hat{H}_c^2(t)\right) = 1 .
  \label{cond2}
 \ee

 \item[(iii)] The background Hamiltonian $\hat{H}_0$ is not energetically dominant, i.e.,
 \be
  {\rm tr}\! \left(\hat{H}_0^2\right) < {\rm tr}\!\left(\hat{H}_c^2(t)\right) .
  \label{cond3}
 \ee
\end{itemize}
These features thus define the quantum counterpart of the classical Zermelo navigation
problem \cite{zermelo31,caratheodory}.

Moved by the possibility to apply the above ideas to more general quantum problems,
here we present a further development of the quantum Zermelo navigation problem, which provides 
a protocol that can be easily adapted to different physical scenarios.
In this regard, we have focused on a series of guidelines, which stress the physics behind
this approach with a limited abstract conceptualization of the problem.
That is, we have tried to answer questions such as whether it is possible to
build a control Hamiltonian without entering too many formal aspects but just well-known theory.
And, if so, we also wanted to know how it looks like and whether it works optimally.
Interestingly, by proceeding this way, we have been able to reach a general form
for the condition specified by \eqref{cond2}, where the left-hand side equals a general
constant $k$, which, in turn, is related to the minimum time necessary to take the
system from the initial to the final quantum state that we wish, circumventing the
unwanted effects of the bare Hamiltonian.
Accordingly, a general protocol is presented to determine $\hat{H}_c$, which we have
tested here with a series of well-known quantum systems, such as the harmonic oscillator,
entanglement swapping with Bell states, or spin-flip in a Heisenberg dimer.
It is worth stressing that, in all cases, although the least time is going to depend
on the system Hamiltonian, the condition itself is totally independent of it, which
we associate with the fact that the evolution of the quantum state keeps a one-to-one
analogy with the geometrical evolution along a meridian joining both states on the
Bloch's sphere, as already pointed out by Brody {\it et al.}~\cite{brogib15}.

The work is organized as follows.
The theory is presented, developed and discussed in the next section.
To be self-contained, a brief account on the classical Zermelo problem as well
as on the Brody and Meier approach is provided, which serves to contextualize the
present work.
Afterwards, in Sec.~\ref{sec2} we present our approach, which also includes a discussion
on the adiabaticity of the solution of the quantum Zermelo problem.
In Sec.~\ref{sec3} we develop the applications mentioned above, showing how the
least-time condition arises in each case.
Finally, a series of concluding remarks are exposed in Sec.~\ref{sec4}.

%%%%%%%%%%%%%%%%%%%%%%%%%%%%%%%%%%%%%%%%%%%%%%%%%%%%%%%%%%%%%%%%%%%%%%%
%%%%%%%%%%%%%%%%%%%%%%%%%%%%%%%%%%%%%%%%%%%%%%%%%%%%%%%%%%%%%%%%%%%%%%%

\section{Theory}
\label{sec2}

%%%%%%%%%%%%%%%%%%%%%%%%%%%%%%%%%%%%%%%%%%%%%%%%%%%%%%%%%%%%%%%%%%%%%%%

\subsection{Classical Zermelo problem}
\label{sec21}

The classical Zermelo navigation problem can be stated as follows.
Given the actual position of a ship, ${\bf x}^\top=(x_1, x_2)$, on the surface of an
unlimited sea and undergoing the local action of a current and/or wind, characterized by a
position-dependent vector field, ${\bf w}^\top({\bf x})=[w_1({\bf x}), w_2({\bf x})]$, one expects
to find the optimal control velocity, ${\bf v}^\top = v\hat{\bf  u}^\top=v(u_1,u_2)$, that should constantly act
on the ship so that it reaches its destination in the least time.
Here, $\hat{\bf u}$ is a unit vector in the direction of ${\bf v}$ and $v$ denotes its modulus.

As it was noticed by Zermelo \cite{zermelo31} and Carath\'{e}odory \cite{caratheodory} in the early 1930's,
the solution to this problem can be obtained by constructing the
geometrical form of the indicatrix that allows one to obtain the Hamiltonian function and, from
it, all extremal curves of the problem.
Accordingly, the absolute velocity
of the ship, namely, ${\bf v} = \dot{\bf x}/F$, must satisfy the equation
\begin{equation}
 \frac{\dot{\bf x}}{F}-{\bf w}({\bf x})=v{\bf u} ,
 \label{eq:1}
\end{equation}
where $\dot{\bf x}$ is the derivative of the coordinates with respect to an arbitrary
evolution parameter.
The time employed by the ship in its full journey is calculated from the
integral of the $F$ function with respect to the arbitrary parameter along the extremal curve.
Hence, $F$ becomes the basic function of this variational problem.

Equation~(\ref{eq:1}) allows us to determine the $F$ function as a positive root of the equation
\begin{equation}
 \left[ \frac{\dot{\bf x}}{F}-{\bf w}({\bf x}) \right]^\top
 \left[ \frac{\dot{\bf x}}{F}-{\bf w}({\bf x}) \right] = v^2 ,
 \label{eq:2}
\end{equation}
whenever such a root exists.
Equation~(\ref{eq:2}) is the indicatrix of the classical Zermelo navigation
problem, which describes a circle of radius $v$ with center at ${\bf w}({\bf x})$.
The set of points satisfying the circle condition corresponds to the end points
of the vector ${\dot{\bf x}}/F$.
As seen below, in the quantum analog for this problem, Brody and Meier \cite{brody15}
found the solution by determining the geodesics of the Randers metric derived
from the form of the $F$ function.

%%%%%%%%%%%%%%%%%%%%%%%%%%%%%%%%%%%%%%%%%%%%%%%%%%%%%%%%%%%%%%%%%%%%%%%

\subsection{Quantum Zermelo approach}
\label{sec22}

While searching for a quantum speed limit to quantum information processing, Russell and Stepney found a tight connection between such processes and the classical Zermelo navigation problem \cite{stepney:PRA:2014}, later on extended to a method to determine optimal times involved in the implementation of quantum gates \cite{stepney:JPA:2015}. Shortly after, Brody {\it et al.}~\cite{brogib15} also reached a similar conclusion, namely, that there is a direct quantum counterpart for the Zermelo navigation problem.
To this end, consider some initial and final quantum states, $|\psi_i\rangle$ and
$|\psi_f\rangle$, respectively, for a given physical system, which is being acted upon by a
time-independent background Hamiltonian $\hat{H}_0$.
The quantum Zermelo problem consists in finding a control Hamiltonian, $\hat{H}_c(t)$,
such that the total Hamiltonian (\ref{hamilton0}) describes a unitary transformation leading from $|\psi_i\rangle$ to $|\psi_f\rangle$ in the least time.
Notice that by invoking to the classical analog, the classical vector field describing
the wind or current corresponds, in the quantum counterpart, to the unitary operator
generated by $\hat{H}_0$.
Furthermore, in this quantum problem, it is assumed that the energy associated with the
transformation from $|\psi_i\rangle$ to $|\psi_f\rangle$ is not only limited but it also has to be totally
consumed at the end of the process.
Thus the speed evolution generated by the control Hamiltonian, $\hat{H}_c(t)$, is
related to the energy variance of $\hat{H}_c(t)$, according to the Anandan-Aharonov
relation \cite{anan90}.
Over the full process, the speed evolution takes the
maximum attainable value and  it is  fixed.

Based on such constraints, one aims to build
an optimal unitary transformation that satisfies them all.
Accordingly, consider the time evolution of the unitary operator, $\hat{U}(t,t_i)$,
governed by the Schr\"odinger equation, which in the Heisenberg representation reads as
\ba
 i\frac{d\hat{U}(t,t_i)}{dt} & = & \hat{H}(t)\hat{U}(t,t_i) \nonumber \\
 & = & \left[ \hat{H}_0 + \hat{H}_c(t) \right] \hat{U}(t,t_i) ,
 \label{eq:3}
\ea
with $\hbar = 1$ (in natural units).
The time-evolution operator $\hat{U}(t,t_i)$ is required to satisfy the initial condition
$\hat{U}(t_i,t_i) \equiv \mathbb{I}$ ($\mathbb{I}$ denotes the identity operator) as well as
the unitarity condition $\hat{U}^\dagger(t,t_i) \hat{U}(t,t_i)=\hat{U}(t,t_i)\hat{U}^\dagger(t,t_i) = \mathbb{I}$,
which ensures the norm preservation along the whole evolution.

For simplicity and convenience, considering the total time lasted in the evolution of the system,
$\Delta T = t_f - t_i$, with $t_i \leq t \leq t_f$, and then defining the dimensionless evolution parameter $s = (t-t_i)/\Delta T$,
the time-evolution operator can be recast as $\hat{U}(t,t_i)=\hat{U}_{\Delta T} (s)$ and its time derivative as
$d\hat{U}(t,t_i)/dt = (1/\Delta T)d\hat{U}_{\Delta T}(s)/ds$ \cite{mess99}.
%  [A. Messiah, Quantum Mechanics (Dover, New York, 1999)].
Using the above notation and multiplying Eq.~(\ref{eq:3}) from the right by $\hat{U}_{\Delta T}^\dagger(t,t_i)$ we get
\begin{equation}
 \frac{i}{\Delta T}\frac{d\hat{U}_{\Delta T}(s)}{ds}\hat{U}_{\Delta T}^\dagger(s) - \hat{H}_0 = \hat{H}_c(s),
 \label{eq:4}
\end{equation}
which strongly resembles the classical Eq.~(\ref{eq:1}), with $\Delta T$ playing the role of $F$.

In order to further stress the quantum-classical analogy, Eq.~(\ref{eq:4}) is now multiplied by itself.
Then the trace over the full resulting evolution equation gives rise to the equation
\begin{widetext}
\begin{equation}
 {\rm tr}\! \left( \hat{X}(s)\hat{X}(s) \right) - 2 \Delta T\, {\rm tr}\! \left( \hat{H}_0\hat{X}(s) \right) + (\Delta T)^2\, {\rm tr}\! \left(\hat{H}_0^2\right) =
 (\Delta T)^2\, {\rm tr}\!\left(\hat{H}_c^2(s)\right) = k (\Delta T)^2 ,
 \label{eq:5}
\end{equation}
with
\be
 \hat{X}(s) = i \frac{d\hat{U}_{\Delta T}(s)}{ds} \hat{U}_{\Delta T}^\dagger (s)
\ee
arising from the constraint on the energy resource bound [see condition (ii) above],
and $k$ being an arbitrary constant (in \cite{brody15}, this constant amounts to 1).
Equation~(\ref{eq:5}) can thus be seen as the quantum counterpart of Eq.~(\ref{eq:2}).
Solving for $\Delta T$ \cite{brody15}, we finally find
\begin{equation}
 \Delta T\{\hat{X}(s)\} = \frac{- {\rm tr}\!\left( \hat{X}(s)\hat{H}_0 \right) + \sqrt{\left[{\rm tr}\!\left(\hat{X}(s)\hat{H}_0\right) \right]^2
 + \left[ k - {\rm tr}\!\left(\hat{H}_0^2\right)\right] {\rm tr}\!\left(\hat{X}(s)\hat{X}(s)\right)}}{k-{\rm tr}\!\left(\hat{H}_0^2\right)} ,
 \label{eq:5aa}
\end{equation}
\end{widetext}
which constitutes the so-called Finslerian norm of $\hat{X}(s)$ \cite{bao04,serr09,alde17}.
As it can readily be noticed, the positivity of $\Delta T$ in
Eq.~(\ref{eq:5aa}) is ensured irrespective of the value of ${\rm tr}\!\left(\hat{H}_0^2\right)$ against $k$.
Note that although in \cite{brody15} Brody and Meier considered that $k$ should be larger (with $k = 1$, in their case), as specified by the above condition (iii), later on,
in \cite{brogib15}, they pointed out that the condition can be actually relaxed in the quantum context due to the compactness of the manifold of pure states.

The question now is whether one can approach the same problem from a more physical
viewpoint, that is, from a more familiar quantum formulation, which, in turn, might
serve also to confer more generality to the process.
The answer is affirmative, as we show now by considering notions already existing
within the time-dependent perturbation theory \cite{mess99},
%[A. Messiah, Quantum Mechanics (Dover, New York, 1999)].
which is also closer to treatments typically considered in the theory of open quantum
systems \cite{breuer-bk:2002}.
To see that let us introduce the unitary time-evolution operator $\hat{U}_0(t,t_i)$,
corresponding to $\hat{H}_0$, solution to the equation
\begin{equation}
 i\frac{d\hat{U}_0(t,t_i)}{dt} = \hat{H}_0\hat{U}_0(t,t_i) ,
 \label{eq:5a}
\end{equation}
with the initial condition $\hat{U}_0(t_i,t_i) \equiv \mathbb{I}$.
The solution is well known,
\begin{equation}
 \hat{U}_0(t,t_i) = e^{-i\hat{H}_0(t-t_i)} .
 \label{eq:5b}
\end{equation}
Now, in order to determine $\hat{U}(t,t_i)$, we consider the separable ansatz
\begin{equation}
 \hat{U}(t,t_i) = \hat{U}_0(t,t_i)\hat{U}_c(t,t_i) ,
 \label{eq:6}
\end{equation}
where the time-evolution operator $\hat{U}_c(t,t_i)$ is required to be unitary and satisfying the unitarity condition $ \hat{U}_c^\dagger(t,t_i) \hat{U}_c(t,t_i) = \mathbb{I}$.
This constraint, in turn, implies that $\hat{U}(t,t_i)$ also satisfies the unitarity condition, as it
can easily be shown.

In order to determine $\hat{U}_c(t,t_i)$, we now proceed as follows.
First we substitute Eq.~(\ref{eq:6}) into Eq.~(\ref{eq:3}) and then make the $\hat{U}_0^\dagger(t,t_i)$
to act on the left of the resulting expression, which renders the equation
%
%\begin{widetext}
\be
 i\frac{d\hat{U}_c(t,t_i)}{dt}
% & = & \hat{U}_0^\dagger(t,t_i)\hat{H}_c(t)\hat{U}_0(t,t_i)\hat{U}_c(t,t_i)
%  + \hat{U}_0^\dagger(t,t_i)\bigg (\hat{H}_0\hat{U}_0(t,t_i)-i\frac{d\hat{U}_0(t,t_i)}{dt}\bigg )\hat{U}_c(t,t_i) \nonumber \\
 = \hat{U}_0^\dagger(t,t_i)\hat{H}_c(t)\hat{U}_0(t,t_i)\hat{U}_c(t,t_i) ,
\label{eq:7}
\ee
%\end{widetext}
%
with initial condition $\hat{U}_c(t_i,t_i) \equiv \mathbb{I}$, and where we have made use of
Eq.~(\ref{eq:5a}) to simplify it.
Now, as it can be seen on the right-hand side, $\hat{U}_c$ is acted upon by the control Hamiltonian operator
in the interaction picture \cite{schiff-bk},
\ba
 \hat{H}'_c(t) & = &\hat{U}_0^\dagger(t,t_i)\hat{H}_c(t)\hat{U}_0(t,t_i) \nonumber \\
 & = & e^{i\hat{H}_0(t-t_i)} \hat{H}_c(t) e^{-i\hat{H}_0(t-t_i)} .
 \label{eq:9}
\ea
Since the control Hamiltonian $\hat{H}_c$ is to be determined, we can make a guess on the
particular functional form for its interaction picture, namely that $\hat{H}'_c$ corresponds
to $\hat{H}_c$ at $t_i$, so that it becomes time independent.
Although this might look counterintuitive, if we recall the picture provided by Brody {\it et al.}~\cite{brogib15}
of the evolution along the Bloch sphere when going from one state to the other, the above condition (\ref{eq:9}),
with $\hat{H}'_c(t) = \hat{H}_c(t_i)$, can be considered to be equivalent to continuing the journey on the back of
the sphere, from the final state to the initial one.
That is, the condition for a proper control requires evolution along a meridian, and no other curve, in order to
ensure the equivalence of the two journeys.

Therefore, even if the above condition seems to be a strong constraint, still it is rather reasonable
and convenient, since it allows us to recast Eq.~(\ref{eq:9}) with a functional form analogous to that found
for the time-evolution operator associated with the bare Hamiltonian, Eq.~(\ref{eq:5a}), i.e.,
\begin{equation}
 i\frac{d\hat{U}_c(t,t_i)}{dt}=\hat{H}_c(t_i)\hat{U}_c(t,t_i) ,
 \label{eq:8}
\end{equation}
with the solution
 \begin{equation}
  \hat{U}_c(t,t_i) = e^{-i\hat{H}_c(t_i)(t-t_i)} .
  \label{eq:9a}
  \end{equation}
Accordingly, the full Hamiltonian for the quantum Zermelo problem acquires the final form
\begin{equation}
 \hat{H}(t) = \hat{H}_0 + e^{-i\hat{H}_0(t-t_i)} \hat{H}_c(t_i) e^{i\hat{H}_0(t-t_i)} ,
 \label{eq:10}
\end{equation}
which corresponds to Eq.~(1) in \cite{brody15}.

Next, let us see some properties that follow from the above relationship between
$\hat{H}_c(t)$ and $\hat{H}_c(t_i)$.
Consider the relation
\begin{equation}
 \hat{H}_c(t) = e^{-i\hat{H}_0(t-t_i)} \hat{H}_c(t_i) e^{i\hat{H}_0(t-t_i)} .
 \label{eq:11}
\end{equation}
It readily follows that if ${\rm tr}\!\left(\hat {H}_c^2 (t_i)\right)$ is constant,
then the same holds for ${\rm tr}\!\left(\hat {H}_c^2(t)\right)$, since
\begin{equation}
 {\rm tr}\! \left(\hat{H}_c^2(t)\right) = {\rm tr}\!\left(\hat{H}_c^2(t_i)\right) = k ,
\label{eq:12}
\end{equation}
which is satisfied at any time $t$.
Thus, according to (\ref{eq:12}), $d{\rm tr}\!\left(\hat {H}_c^2\right)/dt = 0$ also
at any time.
Now, differentiation of Eq.~(\ref{eq:11}) with respect to $t$ leads to
\begin{equation}
 \frac{d\hat{H}_c(t)}{dt} = - i \left[\hat{H}_0, \hat{H}_c(t) \right] ,
\label{eq:13}
\end{equation}
which is a solution to the variational problem, $\delta \int_0^1 \left[ \Delta T \{\hat{X}(s)\} \right]^2 ds = 0$,
with $\Delta T \{\hat{X}(s)\}$ the same as given in Eq.~(\ref{eq:5aa}) and first derived by
Brody and Meier \cite{brody15}.
%[D.C. Brody, and D. M. Meier, Solution to the quantum
%Zermelo Navigation  problem, Phys. Rev. Lett. 114, 100502 (2015)].
Equation~(\ref{eq:13}) gives the coadjoint motion and hence it should be solved
together with Eq.~(\ref{eq:10}).
Besides, from Eq.~(\ref{eq:13}) we also find that ${\rm tr}\!\left(d\hat{H}_c(t)/dt\right) = 0$
and $d{\rm tr}\!\left(\hat {H}_c^2\right)/dt = 2{\rm tr}\!\left(\hat{H}_c(t)d\hat{H}_c(t)/dt\right)=0$
by using cyclic permutation when tracing.
Physically, these vanishing values imply that the ``velocity'' of the transition process remains
constant during the whole process, as it is assumed in the problem by definition.

From the above formulation, it is now clear that Eq.~(\ref{eq:13}) together with Eq.~(\ref{eq:3}),
with $\hat{H}_c(t)$ as given by (\ref{eq:11}), and $\hat{U}(t,t_i)$ computed from (\ref{eq:6}),
(\ref{eq:5b}), and (\ref{eq:9a}), provide the fundamental solution to the quantum Zermelo
problem \cite{brody15,brogib15}.
Furthermore, we have seen that the condition ${\rm tr}\!\left(\hat{H}_c^2(t_i)\right) = k$
arises as a consequence of Eqs.~(\ref{eq:11}) and (\ref{eq:13}) \cite{carli06,carli07} and generalizes the result in Ref.~\cite{brody15}.

%%%%%%%%%%%%%%%%%%%%%%%%%%%%%%%%%%%%%%%%%%%%%%%%%%%%%%%%%%%%%%%%%%%%%%%

\subsection{Transition between two specific quantum states}
\label{sec23}

According to the above results, time optimization in the quantum Zermelo approach
is fully determined by the construction of the control Hamiltonian $\hat{H}_c(t_i)$
provided the bound condition ${\rm tr}(\hat{H}_c^2(t_i)) = k$ is satisfied,
since both the bare Hamiltonian $\hat{H}_0$ and the initial and final states, $|\psi_i\rangle$
and $|\psi_f\rangle$, are given.
In order to understand the dynamical transition from $|\psi_i\rangle$ and $|\psi_f\rangle$,
and hence to introduce a protocol to optimize the time lasting in such a transition, let us
consider the state reached by $|\psi_i\rangle$ after a time $t$ under free evolution, i.e.,
under the action of the bare background Hamiltonian.
The wave function of this state is given by
\be
 |\psi(t)\rangle = \hat{U}(t,t_i) |\psi_i\rangle ,
 \label{barestate}
\ee
where $|\psi_i\rangle = |\psi(t_i)\rangle$.
Taking into account Eq.~(\ref{eq:6}), we can introduce the intermediate state
\begin{equation}
 |\psi'(t)\rangle \equiv \hat{U}_0^\dagger(t,t_i) |\psi(t)\rangle = \hat{U}_c(t,t_i) |\psi_i \rangle .
 \label{eq:14}
\end{equation}
Differentiating this state and its complex conjugate partner with respect to time,
and then substituting the corresponding results in Eq.~(\ref{eq:8}) (and the corresponding
complex conjugate equation) leads to
\begin{subequations}
\ba
 i\frac{d|\psi'(t)\rangle}{dt} & = & \hat{H}_c(t_i) |\psi'(t)\rangle ,
 \label{eq:15a}  \\
 -i\frac{d \langle \psi'(t)|}{dt} & = & \langle \psi'(t)| \hat{H}_c(t_i) .
 \label{eq:15b} \
\ea
\label{eq:15}
\end {subequations}
Now if $|\psi_i\rangle$ is normalized, then $|\psi'(t)\rangle$ is also normalized, as
it can readily be inferred from (\ref{eq:14}).
Moreover, if we assume that the control Hamiltonian generates a state vector that is
orthogonal to the original one (in compliance with the fact that it has to counterbalance
the effect of the ``blowing wind'' accounted for with the bare Hamiltonian), then from (\ref{eq:15})
we have
\be
 \langle \psi'(t)| \frac{d|\psi'(t)\rangle}{dt} = \frac{d\langle \psi'(t)|}{dt} |\psi'(t)\rangle = 0 .
 \label{orthocond}
\ee
In order to satisfy both conditions, normalization and orthogonality, also from (\ref{eq:15})
we notice that $\hat{H}_c(t_i)$ has to display the following functional form \cite{uzd12}:
% [R. Uzdin, U. G\"{u}nther, S. Rahav, and N. Moiseyev, Time-dependent Hamiltonians with 100\% evolution
% speed efficiency J. Phys. A: Math. Theor. 45, 415304 (2012).]
%
\begin{equation}
 \hat{H}_c(t_i) = i \left[ \frac{d|\psi'(t)\rangle}{dt} \langle \psi'(t)| - |\psi'(t)\rangle \frac{d\langle \psi'(t)|}{dt} \right] ,
 \label{eq:16}
\end{equation}
where the right-hand side shows an explicit dependence on time, although the Hamiltonian is time independent.
Rather than an inconsistency, this is just an effect associated with the fact that this Hamiltonian has
to counterbalance at every time the effect produced by $\hat{H}_0$, although the net
action is time independent, as will be shown below.

Notice that the conditions on $|\psi'(t)\rangle$ and its time derivative imply that $\hat{H}_c(t_i)$
is traceless, i.e., ${\rm tr}\! \left(\hat{H}_c(t_i)\right) = 0$.
Moreover, since the variance of the energy is related to the speed of the quantum evolution \cite{anan90},
it can be shown that the orthogonality condition (\ref{orthocond}) ensures the maximum speed evolution
condition for the control Hamiltonian, since it makes the variance of this Hamiltonian, given by the
expression
\ba
 \left(\Delta \hat{H}_c (t_i) \right)^2 & = & \langle \psi '(t)| \hat{H}_c^2(t_i) |\psi'(t)\rangle \nonumber \\
 & & - \left( \langle\psi'(t)| \hat{H}_c(t_i)|\psi'(t)\rangle \right)^2 \nonumber \\
 & = & \frac{d\langle \psi'(t)|}{dt} \left( \mathbb{I} - |\psi'(t)\rangle \langle\psi'(t)| \right) \frac{d|\psi'(t)\rangle}{dt} \nonumber \\
 & = & \left\Vert \frac{d\psi'(t)}{dt} \right\Vert^2 ,
 \label{varianceHc}
\ea
to reach its maximum value.
Actually, we have that
\be
 2 \left(\Delta \hat{H}_c (t_i) \right)^2 = {\rm tr}({\hat H}_c^2(t_i)) = k ,
 \label{traceHc2}
\ee
which is a consequence of the fact that the control Hamiltonian is traceless \cite{carli07,brogib15}.

From Eqs.~(\ref{varianceHc}) and (\ref{traceHc2}), we find the following relation:
\be
 \left\Vert \frac{d\psi'(t)}{dt} \right\Vert^2 = \frac{k}{2} .
\ee
At any time, this relation is satisfied by the ansatz
\ba
 |\psi'(t)\rangle & = & \cos \left[\sqrt{k/2}(t-t_i)\right] |\psi'(t_i)\rangle \nonumber \\
 & & + \frac{\sin \left[\sqrt{k/2}(t-t_i)\right]}{\sqrt{k/2}}\frac{d|\psi'(t_i)\rangle}{dt} ,
 \label{eq:17a}
\ea
with time derivative given by
\ba
 \frac{d|\psi'(t)\rangle}{dt} & = & -\sqrt{k/2}\sin \left[\sqrt{k/2}(t-t_i)\right]|\psi'(t_i)\rangle \nonumber \\
 & & + \cos\left[\sqrt{k/2}(t-t_i)\right]\frac{d|\psi'(t_i)\rangle}{dt} .
 \label{eq:17b}
\ea
This ansatz, in turn, satisfies the above normalization and orthogonality conditions.
Notice here that the expression $d|\psi'(t_i)\rangle/dt$ has to be understood as the
time derivative of $|\psi'(t)\rangle$ evaluated at $t=t_i$.

In order to further simplify the approach, the above expressions \eqref{eq:17a} and \eqref{eq:17b}, in terms
of the general time-evolved state vector $|\psi'(t)\rangle$, can be recast in terms of the
initial and final-state vectors, $|\psi_i\rangle$ and $|\psi_f\rangle$, thus also providing an
simpler functional form for the control Hamiltonian \eqref{eq:16}.
To this end, notice that by virtue of Eq.~(\ref{eq:14}), at $t_i$ we have
$|\psi'(t_i)\rangle = |\psi_i\rangle$.
Similarly, at $t_f$ we find
\begin{equation}
 \hat{U}_0^\dagger (t_f,t_i) |\psi_f\rangle = \hat{U}_c(t_f,t_i) |\psi_i\rangle
  = |\psi'(t_f)\rangle = |\psi'_f\rangle .
 \label{eq:17c}
\end{equation}
In order to remove any common support between $|\psi'_f\rangle$ and $|\psi_i\rangle$,
we need to find the orthonormal form for the former, which is obtained by applying
a Gram-Schmidt orthogonalizing process.
Accordingly, the orthonormal form is found to be
\ba
 |\bar{\psi}'_f\rangle
% & = & \left( \mathbb{I} - |\psi_i\rangle \langle \psi_i| \right) \hat{U}_0^\dagger(t_f,t_i)|\psi_f\rangle \nonumber \\
 & = & \left( \mathbb{I} - |\psi_i\rangle \langle \psi_i| \right) |\psi'_f\rangle \nonumber \\
 & = & \frac{\sin \left(\sqrt{k/2} \Delta T \right)}{\sqrt{k/2}}\frac{d|\psi'(t_i)\rangle}{dt} ,
 \label{eq:18}
\ea
where Eq.~(\ref{eq:17a}) has been used, with $t=t_f$.
Next, we normalize $|\bar{\psi}'_f\rangle$:
\ba
 |\bar{\bar{\psi}}'_f\rangle & = & \frac{1}{\left\Vert \bar{\psi}'_f \right\Vert}\ |\bar{\psi}'_f\rangle \nonumber \\
 & = & \frac{\left( \mathbb{I} - |\psi_i\rangle \langle\psi_i| \right) \hat{U}_0^\dagger(t_f,t_i) |\psi_f\rangle}
 {\sqrt{1 - \left\arrowvert \langle \psi_f|\hat{U}_0(t_f,t_i)|\psi_i\rangle \right\arrowvert^2}} \nonumber \\
 & = & \frac{1}{\sqrt{k/2}}\frac{d|\psi'(t_i)\rangle}{dt} .
\label{eq:19}
\ea
It can be noticed from Eq.~(\ref{eq:19}) that the calculation of $|\bar{\bar{\psi}}'_f\rangle$ only includes $|\psi_i\rangle$, $|\psi_f\rangle$, $\hat{H}_0$, and the time interval $\Delta T$.

The ansatz~(\ref{eq:17a}) and its time derivative, Eq.~(\ref{eq:17b}), can now be recast in terms of the orthonormal state vectors $|\psi_i\rangle$ and
$|\bar{\bar{\psi}}'_f\rangle$, which read as
\begin{subequations}
\ba
 |\psi'(t)\rangle & = & \cos \left[\sqrt{k/2}(t-t_i)\right] |\psi_i\rangle \nonumber \\
 & & + \sin \left[\sqrt{k/2}(t-t_i)\right] |\bar{\bar{\psi}}'_f\rangle ,
 \label{eq:20a} \\
 \frac{d|\psi'(t)\rangle}{dt} & = & -\sqrt{k/2} \sin\left( \sqrt{k/2}(t-t_i) \right) |\psi_i\rangle \nonumber \\
 & & + \sqrt{k/2} \cos \left(\sqrt{k/2}(t-t_i)\right) |\bar{\bar{\psi}}'_f\rangle ,
 \label{eq:20b}
\ea
\label{eq:20}
\end{subequations}
respectively.

In order to finally obtain the functional form of the control Hamiltonian, we substitute
Eqs.~(\ref{eq:20}) into (\ref{eq:16}), leading to
\begin{equation}
 \hat{H}_c(t_i) = i \sqrt{k/2} \left[ |\bar{\bar{\psi}}'_f\rangle \langle\psi_i|
  - |\psi_i\rangle \langle \bar{\bar{\psi}}'_f| \right] ,
 \label{eq:21}
\end{equation}
which is the initial optimal control Hamiltonian.
With the aid of Eq.~(\ref{eq:20a}), in the case $t = t_f$, $|\bar{\bar{\psi}}'_f\rangle$ can
be recast in terms of $|\psi'_f\rangle = |\psi'(t_f)\rangle$.
If the corresponding expression is then substituted into Eq.~(\ref{eq:21}), we shall obtain
\begin{equation}
 \hat{H}_c(t_i) = i \frac{\sqrt{k/2}}{\sin\left( \sqrt{k/2} \Delta T \right)}
\left[ |\psi'_f\rangle \langle\psi_i| - |\psi_i\rangle \langle\psi'_f| \right] ,
\label{eq:21iden}
\end{equation}
which is time independent, as it was stressed above.
This is precisely the expression reported by Brody {\it et al.}~\cite{brogib15} for $\hat{H}_c$ in the particular case $k = 1$.
It can be shown now that the variance of $\hat{H}_c(t_i)$ for any $|\psi'(t)\rangle$ effectively
remains constant in time, that is,
$\Delta \hat{H}_c(t_i) = \sqrt{ \langle \psi'(t)| \hat{H}_c^2(t_i)|\psi'(t)\rangle} =\sqrt{k/2}$.

The expression of the optimal control Hamiltonian (\ref{eq:21}) can be written in diagonal form as
\begin{widetext}
\begin{equation}
 \hat{H}_c(t_i) = \frac{1}{\sqrt{2}} \begin{pmatrix} |\psi_i\rangle - i |\bar{\bar{\psi}}'_f\rangle , & |\psi_i\rangle + i |\bar{\bar{\psi}}'_f\rangle \end{pmatrix}
 \begin{pmatrix} -\sqrt{k/2} & 0 \\ 0 & \sqrt{k/2} \end{pmatrix}
 \frac{1}{\sqrt{2}} \begin{pmatrix} \langle\psi_i| + i\langle\bar{\bar{\psi}}'_f| \\ \langle\psi_i| - i \langle \bar{\bar{\psi}}'_f| \end{pmatrix} .
\label{eq:21a}
\end{equation}
With this expression at hand, Eq.~(\ref{eq:9a}) takes the explicit form
\be
 \hat{U}_c(t,t_i)
%  =\exp(-i\hat{H}_c(t_i)(t-t_i)) =
 \frac{1}{\sqrt{2}} \begin{pmatrix} |\psi_i\rangle - i |\bar{\bar{\psi}}'_f\rangle , & |\psi_i\rangle + i |\bar{\bar{\psi}}'_f\rangle \end{pmatrix}
 \begin{pmatrix} e^{i\sqrt{k/2}(t-t_i)} & 0 \\ 0 & e^{-i\sqrt{k/2}(t-t_i)} \end{pmatrix}
 \frac{1}{\sqrt{2}} \begin{pmatrix} \langle\psi_i| + i\langle\bar{\bar{\psi}}'_f| \\ \langle\psi_i| - i \langle \bar{\bar{\psi}}'_f| \end{pmatrix} .
 \label{eq:21b}
\ee
\end{widetext}
The time interval $\Delta T$ is then evaluated by considering the transformation
indicated in Eq.~(\ref{eq:17c}), i.e., the time-operator $\hat{U}_c(t_f,t_i)$ that takes $|\psi_i\rangle$
to $|\psi'_f\rangle$ in the shortest time.
As it can be noticed in the above expression,
the initial state vector $|\psi_i\rangle$ and the intermediate one
$|\psi'_f\rangle$ are directly related by a phase factor, with its argument providing a
measure of their angular distance.
This angle arises from the overlapping integral
between the final states led by the total Hamiltonian $H$, on the one hand, and the bare
Hamiltonian $H_0$, on the other hand, and reads as
\ba
 \phi & \equiv & \cos^{-1} \left( \left\arrowvert \langle \psi_i| \hat{U}_0^\dagger(t_f,t_i) |\psi_f\rangle \right\arrowvert \right) \nonumber \\
  & = & \Delta T \Delta\hat{H}_c(t_i) \nonumber \\
  & = & \sqrt{k/2} \Delta T ,
 \label{eq:22}
\ea
in compliance with what is stated in the literature on the geometry of the state vector evolution \cite{anan90,vaid92}.

From the above discussion, we then extract as a conclusion that in order to make the state vector to evolve
in the shortest time from $|\psi_i\rangle$ to $|\psi_f\rangle$, when there is the influence of a background
Hamiltonian $\hat{H}_0$, we need to determine the time-optimal unitary transformation, $\hat{U}(t_f,t_i)$,
which includes the following steps:
\begin{enumerate}
 \item Given $\hat{H}_0$, $|\psi_i\rangle$, $|\psi_f\rangle$, and $k$ (the energy bound), compute the time
  interval $\Delta T$ recursively by means of Eq.~(\ref{eq:22}), and the unitary transformation
  $\hat{U}_0(t_f,t_i)$ by means of Eq.~(\ref{eq:5b}).

 \item With $|\psi_i\rangle$, $|\psi_f\rangle$, and $\hat{U}_0(t_f,t_i)$, compute $|\bar{\bar{\psi}}'_f\rangle$
  by means of Eq.~(\ref{eq:19}).

 \item Compute $\hat{U}_c(t_f,t_i)$ using $|\psi_i\rangle$, $|\bar{\bar{\psi}}'_f\rangle$, $k$, and $\Delta T$, according to Eq.~(\ref{eq:21b}).

 \item Using $\hat{U}_0(t_f,t_i)$ and $\hat{U}_c(t_f,t_i)$, compute the time-optimal quantum Zermelo unitary
  transformation, $\hat{U}(t_f,t_i)$, according to Eq.~(\ref{eq:6}).
\end{enumerate}
This protocol will ensure that the unitary transformation $\hat{U}(t_f,t_i)$ transforms $|\psi_i\rangle$
into $|\psi_f\rangle$ in the least time.

Before we move to the practical examples, let us briefly comment on the adiabaticity of the quantum Zermelo Hamiltonian.
At a fundamental level, it would be interesting to establish a direct connection between the quantum Zermelo navigation problem and optimal control theory~\cite{glaser:EPJD:2015}, where adiabaticity plays a key
role as a technique to reach the target by its continuous monitoring over long times.
As it is shown below, the Zermelo navigation problem belongs, by construction, to a different type of optimization scheme, reminiscence of optimal control
schemes related to the classical isoperimetric problem \cite{sargent:JCAM:2000}, and therefore adiabaticity does not arise naturally.

As it can be noticed from the above discussion, Eq.~(\ref{eq:6}) is the solution of Eq.~(\ref{eq:3}),
where $\hat{U}_0(t,t_i)$ and $\hat{U}_c(t,t_i)$ are given by Eqs.~(\ref{eq:5b}) and (\ref{eq:9a}),
respectively.
Therefore, unless Eq.~\eqref{eq:13} is zero, then according to the Baker-Campbell-Hausdorff formula,
\be
 \hat{U}(t,t_i) \neq e^{-i [\hat{H}_0+\hat{H}_c(t_i)](t-t_i)}.
\ee
This result implies that the solution to the quantum Zermelo navigation problem does not define an adiabatic process in general. To see this, let $\left \{ \phi_j(t_i)  \right \}_{j=1}^{N}$ denote the orthonormal set of eigenfunctions
of $\hat{H}(t_i)$ and $\left \{ h_j  \right \}_{j=1}^{N}$ the corresponding set of eigenvalues,
with $N$ being the dimension of the space.
The eigenvalues $h_j$ are time independent, since ${\rm tr}\! \left(d\hat{H}(t)/dt\right) = {\rm tr}\! \left(d\hat{H}_c(t)/dt\right)=0$,
as proven above.
Now, the action of $\hat{U}(t,t_i)$ on an eigenfunction $|\phi_k(t_i)\rangle$ with eigenvalue $h_k$ gives
\be
 U(t,t_i) \phi_j(t_i) \neq \phi_j(t),
\ee
and therefore we conclude that $\langle \psi(t_i)|\phi_k(t_i)\rangle  \neq  \langle \psi(t)| \phi_k(t)\rangle$. That is, if the system is initially represented by the wave function $|\psi(t_i)\rangle = \sum_k c_k(t_i) |\phi_k(t_i)\rangle$ where $c_k(t_i) = \langle \phi_k(t_i) | \psi (t_i)\rangle$, then the probability that the system is in the stationary state $|\phi_k(t)\rangle$ at any time $t$ is not constant, i.e., $\frac{d}{dt}|\langle
\psi(t)|\phi_k(t)\rangle|^2 \neq 0$. This result proves that the dynamical transformation governed by Eq.~(\ref{eq:3})
taking $\hat{U}(t,t_i)$ as that given in Eq.~(\ref{eq:6}) does not
define an adiabatic process~\cite{bornfock1928,kato1949,mess99}.

%%%%%%%%%%%%%%%%%%%%%%%%%%%%%%%%%%%%%%%%%%%%%%%%%%%%%%%%%%%%%%%%%%%%%%%
%%%%%%%%%%%%%%%%%%%%%%%%%%%%%%%%%%%%%%%%%%%%%%%%%%%%%%%%%%%%%%%%%%%%%%%

\section{Applications}
\label{sec3}

%%%%%%%%%%%%%%%%%%%%%%%%%%%%%%%%%%%%%%%%%%%%%%%%%%%%%%%%%%%%%%%%%%%%%%%

\subsection{Harmonic oscillator}
\label{sec31}

We shall start the application of the above-described protocol with the
paradigmatic harmonic oscillator acted on by an external field.
In particular, we are going to consider a two-level transition, which, for
simplicity, is going to be considered to be the ground and the first excited
state, which can be denoted as
\be
 |0\rangle = \begin{pmatrix}1\\0\end{pmatrix} , \quad
 |1\rangle = \begin{pmatrix}0\\1\end{pmatrix} ,
 \label{qbits}
\ee
respectively.
Let us consider the transition from the ground state to the excited state,
so that $|\psi_i\rangle = |0\rangle$ and $|\psi_f\rangle = |1\rangle$.
Of course, these states are under the action of the harmonic oscillator
Hamiltonian,
\begin{equation}
 \hat{H}_0 = \hbar\omega\left( \hat{a}^\dagger \hat{a} + \frac{1}{2}\right) ,
 \label{harmonic}
\end{equation}
with frequency $\omega$.
So, in principle, if the system is isolated, their only time dependence is in terms of a phase
factor; if they form a linear superposition, there will be a periodic transition from one
to the other, with frequency equal to the oscillator frequency, since
$\epsilon_f - \epsilon_i = \hbar \omega$, where $\epsilon_f$ and $\epsilon_i$ are the energies
of states $|1\rangle$ and $|0\rangle$, respectively.
Besides, it is interesting to note that the creation and annihilation operators included
in (\ref{harmonic}), in terms of the states (\ref{qbits}), can be written as
\begin{equation}
 \hat{a} = |0\rangle\langle 1| = \begin{pmatrix} 0 & 1 \\ 0 & 0 \end{pmatrix} , \quad
 \hat{a}^\dagger = |1\rangle\langle 0| = \begin{pmatrix} 0 & 0 \\ 1 & 0 \end{pmatrix} .
\end{equation}

The minimum control time is $\Delta T = \pi/\sqrt{2k}$, and the control Hamiltonian
in (\ref{eq:21iden}) can be written as
\ba
 \hat{H}_c(t_i) & = & i\sqrt{\frac{k}{2}} \left[ e^{-3\pi i\hbar\omega/2\sqrt{2k}} \hat{a}^\dagger
  - e^{3\pi i\hbar\omega/2\sqrt{2k}} \hat{a} \right] \nonumber \\
  & = & i\sqrt{\frac{k}{2}}\cos\left(\frac{3\pi\hbar\omega}{2\sqrt{2k}}\right)\left( \hat{a}^\dagger - \hat{a} \right) \nonumber \\
 & & \qquad + \sqrt{\frac{k}{2}}\sin \left(\frac{3\pi\hbar\omega}{2\sqrt{2k}}\right)\left( \hat{a}^\dagger + \hat{a} \right) .
 \label{Hc_qubit}
\ea
In order for this Hamiltonian to be assimilated by a standard external driving force, one needs to check the following condition:
\begin{equation}
 \hat{H}_c(t_i) = -\sqrt{\frac{\hbar}{2\omega}} \left(\hat{a}^\dagger + \hat{a}\right) E_0 ,
\end{equation}
where $E_0$ is the amplitude of the external electric driving field.
A simple inspection allows us to realize that the above equation is fulfilled only if
$\cos(3\pi\hbar\omega/2\sqrt{2k}) = 0$, which leads to the conclusion
\begin{equation}
 \label{k_qubit}
 k = \frac{(3/2\hbar\omega)^2}{2(n + 1/2)^2} = \frac{\epsilon_{f}^2}{2(n + 1/2)^2} ,
\end{equation}
with $n \in \mathbb{Z}$.

Therefore, given a frequency $\omega$, the maximum $k$ is given by
\begin{equation}
 \label{min_k_qubit}
 k = 2\epsilon_{f}^2 ,
\end{equation}
which corresponds to the minimum control time
\begin{equation}\label{min_k_Delta_t}
 \Delta T = \frac{\pi}{2|\epsilon_{f}|} .
\end{equation}

As it will be seen below, these results are in compliance with those for the
Heisenberg spin dimer, thus paving the way to intuitively consider that
there might an underlying common pattern for any quantum system in the form
of $k$ that provides a physical (implementable) control.

%%%%%%%%%%%%%%%%%%%%%%%%%%%%%%%%%%%%%%%%%%%%%%%%%%%%%%%%%%%%%%%%%%%%%%%

\subsection{Entanglement swapping}
\label{sec32}

Let us now consider entanglement swapping with maximally entangled states
of a Bell basis \cite{bell64,avronjmp2007,avronrmp2020}, where the two
entangled qubits are assumed to be spatially distant, a
paradigm with special interest in quantum information and quantum computation
\cite{llo96,nielsen-chuang-bk}.
More specifically, here we consider two spins, $\hat{\bf \sigma}^{(1)}$ and $\hat{\bf \sigma}^{(2)}$, interacting via anisotropic time-independent
$J_j$ couplings, with $(j=x,y,z)$, under the effect of local, uniform and controllable magnetic fields
$B^{(i)}(t)$, with $(i=1,2)$, pointing along the $z$ direction.
Thus, we choose to consider the following two-qubit Heisenberg Hamiltonian \cite{carli07},
\begin{equation}
 \hat{H} = -\sum_j J_j \hat{\sigma}_j^{(1)} \hat{\sigma}_j^{(2)} + \sum_{i=1}^2 B^{(i)} \hat{\sigma}_z^{(i)} ,
 \label{eq:23}
\end{equation}
to be the quantum Zermelo Hamiltonian $\hat{H}(t)$.
Here, we use the tensor products $\hat{\sigma}_j^{(1)} = \hat{\sigma}_j \otimes \mathbb{I}$ and
$\hat{\sigma}_j^{(2)} = \mathbb{I} \otimes \hat{\sigma}_j$, with $\mathbb{I}$ being the unit operator of
dimension $2\times2$, and $\hat{\sigma}_j^{(i)}$ the Pauli matrices \cite{mess99}.

A simpler ansatz for $\hat{H}$ was already reported in \cite{li11}, where only a fixed
coupling, $J$, was considered.
Here, we are going to associate the first term in (\ref{eq:23}) with the non-controlled,
time-independent background Hamiltonian, $\hat{H}_0$, and the second term with the time-dependent
control Hamiltonian, $\hat{H}_c(t)$, satisfying the energy resource bound,
${\rm tr}\!\left( \hat{H}_c^2(t) \right)=k$.
In this case, the computational basis set is provided by the following factorizable state vectors:
\begin{subequations}
\label{eq:24}
\ba
 \big|00\rangle & = & \begin{pmatrix}1\\0 \end{pmatrix}
     \otimes\begin{pmatrix}1\\0 \end{pmatrix} = \begin{pmatrix} 1 & 0 & 0 & 0 \end{pmatrix}^\top ,
 \label{eq:24a} \\
 \big|01\rangle & = & \begin{pmatrix}1\\0 \end{pmatrix}
     \otimes\begin{pmatrix}0\\1 \end{pmatrix} = \begin{pmatrix} 0 & 1 & 0 & 0 \end{pmatrix}^\top ,
 \label{eq:24b} \\
 \big|10\rangle & = & \begin{pmatrix}0\\1 \end{pmatrix}
     \otimes\begin{pmatrix}1\\0 \end{pmatrix} = \begin{pmatrix} 0 & 0 & 1 & 0 \end{pmatrix}^\top ,
 \label{eq:24c} \\
 \big|11\rangle & = & \begin{pmatrix}0\\1 \end{pmatrix}
     \otimes\begin{pmatrix}0\\1 \end{pmatrix} = \begin{pmatrix} 0 & 0 & 0 & 1 \end{pmatrix}^\top .
 \label{eq:24d}
\ea
\end{subequations}
In this basis, $\hat{H}_0$ reads as \cite{carli07}
\begin{equation}
 \hat{H}_0 = \begin{pmatrix} -J_z & 0 & 0 & -J_- \\ 0 & J_z & -J_+ & 0 \\
 0 & -J_+ & J_z & 0 \\ -J_- & 0 & 0 & -J_z \end{pmatrix} ,
 \label{eq:25}
\end{equation}
where $J_\pm=J_x \pm J_y$.
The diagonal form for $\hat{H}_0$ is
\ba
 \hat{H}_0 & = & - (J_z + J_-)|\Phi_+\rangle\langle \Phi_+|
                 - (J_z - J_-)|\Phi_-\rangle\langle \Phi_-| \nonumber \\
             & & + (J_z - J_+)|\overline{\Phi}_+\rangle\langle \overline{\Phi}_+|
                 + (J_z + J_+)|\overline{\Phi}_-\rangle\langle \overline{\Phi}_-| , \nonumber \\
 & &
 \label{eq:26}
\ea
which allows us to rearrange the above basis set in terms of the Bell basis of
maximally entangled states, namely,
\begin{subequations}
\ba
 |\Phi_+\rangle & = & \frac{1}{\sqrt{2}} \left( |00\rangle + |11\rangle \right) , \\
 |\Phi_-\rangle & = & \frac{1}{\sqrt{2}} \left( |00\rangle - |11\rangle \right) , \\
 |\overline{\Phi}_+\rangle & = & \frac{1}{\sqrt{2}} \left( |01\rangle + |10\rangle \right) , \\
 |\overline{\Phi}_-\rangle & = & \frac{1}{\sqrt{2}} \left( |01\rangle - |10\rangle \right) .
\ea
 \label{Bellbasis}
\end{subequations}
Now the question is how to reach one of these basis vectors from another of them, for instance,
the $| \psi_f\rangle = |\Phi_-\rangle$ state from the $|\psi_i \rangle = |\Phi_+\rangle$ state, in the shortest time using the
optimal-time Zermelo unitary transformation, Eq.~(\ref{eq:6}).

The first term of the unitary time transformation Eq.~(\ref{eq:6}), namely, $\hat{U}_0(t,t_i)$,
is easily obtained from the spectral decomposition of $\hat{H}_0$ given in Eq.~(\ref{eq:26}):
\ba
 \hat{U}_0(t,t_i) & = & e^{i(J_z+J_-)\Delta t} |\Phi_+\rangle\langle \Phi_+| \nonumber \\
 & & + e^{i(J_z-J_-)\Delta t} |\Phi_-\rangle\langle \Phi_-| \nonumber \\
 & & + e^{-i(J_z-J_+)\Delta t} |\overline{\Phi}_+\rangle\langle \overline{\Phi}_+| \nonumber \\
 & & + e^{-i(J_z+J_+)\Delta t} |\overline{\Phi}_-\rangle\langle \overline{\Phi}_-| ,
\label{eq:27}
\ea
with $\Delta t=t-t_i$.
The calculation of the second term of Eq.~(\ref{eq:6}), $\hat{U}_c(t)$, is a bit more subtle.
As mentioned above, we are interested in the transformation of $|\Phi_+\rangle$ into $|\Phi_-\rangle$
via the unitary transformation $|\Phi_-\rangle = \hat{U}_0(t,t_i)\hat{U}_c(t,t_i)|\Phi_+\rangle$
in the shortest time possible.
As explained above, in the previous section, $\hat{U}_c(t,t_i)$ transforms $|\Phi_+\rangle$ into
$\hat{U}_0^\dagger(t,t_i)|\Phi_-\rangle = |\Phi'_-\rangle$ [see Eq.~(\ref{eq:14})].
Accordingly, in the present case, $|\Phi'_-\rangle = |\Phi_-\rangle \exp[i(J_z-J_-)\Delta t]$,
where we have made use of (\ref{eq:27}).
The intermediate state $|\Phi'_-\rangle$ satisfies the relations $\langle \Phi'_-| \Phi'_-\rangle = 1$
and $\langle \Phi'_-|\Phi_+\rangle = 0$, and hence $\hat{H}_c(t_i)$ will have the functional form
\ba
 \hat{H}_c( t_i) & = & i\sqrt{k/2} \left( |\Phi'_{-m}\rangle\langle\Phi_+| - |\Phi_+\rangle\langle\Phi'_{-m}| \right) \nonumber \\
 & = & i\sqrt{k/2} \left[ e^{i(J_z-J_-)\Delta T} |\Phi_-\rangle\langle\Phi_+| \right. \nonumber \\
 & & \qquad \left. - e^{-i(J_z-J_-)\Delta T} |\Phi_+\rangle\langle\Phi_-| \right] ,
\label{eq:28}
\ea
where as noted before, $\Delta T =t_f -t_i$ is the minimum time interval to be determined, and $|\Phi'_{-m}\rangle = |\Phi'_{-}\rangle$
for $\Delta t=\Delta T$.

The next task consists in transforming the $\hat{H}_c(t)$ form specified in the second term of
Eq.~(\ref{eq:23}) into the $\hat{H}_c(t_i)$ form of Eq.~(\ref{eq:28}), as it was also done in
the case of the harmonic oscillator.
In the basis set (\ref{eq:24}), the control Hamiltonian $\hat{H}_c(t)$ reads as
\begin{equation}
 \hat{H}_c( t) = \begin{pmatrix} B_+ & 0 & 0 & 0 \\ 0 & B_- & 0 & 0 \\
  0 & 0 & -B_- & 0 \\ 0 & 0 & 0 & -B_+ \end{pmatrix} ,
 \label{eq:29}
\end{equation}
where $B_{\pm} = B^{(1)} \pm B^{(2)}$ (the time dependence in $ B^{(1)}$ and $B^{(2)}$ has been
dropped for simplicity).
As it can be noticed, ${\rm tr}\!\left(\hat{H}_c(t)\right) = 0$, but ${\rm tr}\!\left(\hat{H}_c^2(t)\right)$
is not constant in time because the $\hat{H}_c(t)$ form in Eq.~(\ref{eq:29}) does not involve
time unitarity.
Hence, next we have to transform the $\hat{H}_c( t)$ in Eq.~(\ref{eq:29}) into the form given by
Eq.~(\ref{eq:11}) with an appropriate choice of $\hat{H}_c(t_i)$, according to Eq.~(\ref{eq:21}).
The projection of $H_c(t_i)$ given by Eq.~(\ref{eq:21}) onto the subspace spanned by $|\psi_i\rangle$
and $|\bar{\bar{\psi}}'_f\rangle$ results in two vanishing diagonal elements and two off-diagonal
elements with zero real part, where their imaginary part is equal to $\sqrt{k/2}$.
Analogously, we project the $\hat{H}_c(t)$ from Eq.~(\ref{eq:28}) onto the subspace spanned by
$|\Phi_+\rangle$ and $|\Phi'_{-m}\rangle$. In this new representation, we have
\be
 \langle\Phi_+|\hat{H}_c(t)|\Phi_+\rangle = \langle\Phi'_{-m}|\hat{H}_c(t)|\Phi'_{-m}\rangle = 0 ,
\ee
whereas
\ba
 \langle\Phi'_{-m}|\hat{H}_c(t)|\Phi_+\rangle & = & B_+ e^{i(J_z-J_-)\Delta T} \nonumber \\
 & = & B_+ \left[ \cos \left[(J_z - J_-)\Delta T\right] \right. \nonumber \\
 & & + \left. i \sin \left[(J_z-J_-)\Delta T \right] \right] ,
\ea
where, effectively, we notice
\begin{subequations}
\ba
 {\rm Re} \langle\Phi'_{-m}|\hat{H}_c(t)|\Phi_+\rangle & = & B_+ \cos \left[ (J_z-J_-)\Delta T \right] = 0 , \nonumber \\ & & \\
 {\rm Im} \langle\Phi'_{-m}|\hat{H}_c(t)|\Phi_+\rangle & = & B_+ \sin \left[ (J_z-J_-)\Delta T \right] = \sqrt{k/2} . \nonumber \\ & &
\ea
\end{subequations}
On the other hand, from Eq.~(\ref{eq:22}),
\be
 \cos^{-1} \left( \langle\Phi_+|\Phi'_{-m}\rangle \right) = \pi/2 = \Delta T\sqrt{k/2} ,
\ee
which renders
\be
 \Delta T = \frac{\pi}{\sqrt{k/2}} .
\ee
Substituting the value $\Delta T$ into the real part, we have $\sqrt{k/2} = J_z-J_-$, while
if the substitution is made into the imaginary part, then $B_+ = \sqrt{k/2}$, since $B_+ \neq 0$.
Furthermore, the control variable $B_+$ decouples from the others, namely $B_+ = B_{0+} \cos [2(\mu t + \nu)]$,
where $B_{0+}$, $\mu$ and $\nu$ are time-independent constants.
Taking $\mu = \nu = 0$, $B_{0+} = J_z-J_- = \sqrt{k/2}$, and $\Delta T = (\pi/2)(J_z-J_-)^{-1} = (\pi/2)(B_{0+})^{-1}$,
we reach the final form for $\hat{H}_c(t_i)$, which reads as
\ba
 \hat{H}_c( t_i) & = & B_{0+} \left(|\Phi_-\rangle\langle\Phi_+|+|\Phi_+\rangle\langle\Phi_-|\right) \nonumber \\
 & = & B_{0+} \left(|\Psi_+\rangle\langle\Psi_+| - |\Psi_-\rangle\langle\Psi_-| \right) \nonumber \\
 & = & \frac{B_{0+}}{2} \left(\hat{\sigma}_z \otimes \mathbb{I} + \mathbb{I} \otimes \hat{\sigma}_z \right) \nonumber \\
 & = & \frac{B_{0+}}{2} \left(\hat{\sigma}_z^{(1)} + \hat{\sigma}_z^{(2)} \right) ,
 \label{eq:30}
\ea
where $|\Psi_+\rangle = (|\Phi_+\rangle + |\Phi_-\rangle)/\sqrt2$ and
$|\Psi_-\rangle = (|\Phi_+\rangle-|\Phi_-\rangle)/\sqrt2$.
With this, the corresponding unitary transformation is given by
\begin{equation}
 \hat{U}_c(t,t_i) = e^{-iB_{0+}\Delta t}|\Psi_+\rangle\langle\Psi_+|
 + e^{iB_{0+}\Delta t}|\Psi_-\rangle\langle\Psi_-| .
 \label{eq:31}
\end{equation}
Finally, using Eqs.~(\ref{eq:27}) and (\ref{eq:31}), we obtain the time-optimal
quantum Zermelo unitary transformation that leads the Bell basis vector $|\Phi_+\rangle$
into $|\Phi_-\rangle$, namely,
\ba
 |\Phi_-\rangle & = & \hat{U}_z(t,t_i) |\Phi_+\rangle = \hat{U}_0(t,t_i)\hat{U}_c(t,t_i)|\Phi_+\rangle \nonumber \\
 & = & \frac{1}{2} \left[ e^{i(J_z+J_-)\Delta t} |\Phi_+\rangle \left[ e^{-iB_{0+}\Delta t} + e^{iB_{0+}\Delta t} \right] \right. \nonumber \\
 & & \left. + e^{i(J_z-J_-)\Delta t} |\Phi_-\rangle \left[ e^{-iB_{0+}\Delta t} - e^{iB_{0+}\Delta t} \right] \right] , \nonumber \\
 \label{eq:32}
\ea
with $0\leq \Delta t \leq \Delta T$.
As it can be noticed, once the journey is complete, i.e., $\Delta t = \Delta T$,
the Bell state $|\Phi_-\rangle$ is reached.

It is worth noting that in the basis set (\ref{Bellbasis}), the quantum Zermelo Hamiltonian
acquires the form
\begin{widetext}
\ba
 \hat{H}_z(t_i) & = & \hat{H}_0 + \hat{H}_c(t_i) \nonumber \\
 & = & - (J_z + J_-) |\Phi_+\rangle\langle \Phi_+| - (J_z - J_-) |\Phi_-\rangle\langle \Phi_-|
  + (J_z - J_+) |\overline{\Phi}_+\rangle\langle\overline{\Phi}_+|
  + (J_z + J_+) |\overline{\Phi}_-\rangle\langle\overline{\Phi}_-| \nonumber \\
 & & + B_{0+} \left[ |\Phi_-\rangle\langle \Phi_+| + |\Phi_+\rangle\langle \Phi_-| \right] \nonumber \\
 & = & \begin{pmatrix} |\Phi_+\rangle , & |\Phi_-\rangle , & |\overline{\Phi}_+\rangle , & |\overline{\Phi}_-\rangle \end{pmatrix}
 \begin{pmatrix}
   -(J_z+J_-) &    B_{0+}   &       0             &       0       \\
      B_{0+} &  -(J_z-J_-) &       0             &       0        \\
        0         &        0       &  (J_z-J_+)  &       0         \\
        0         &        0       &       0     &   (J_z+J_+) \end{pmatrix}
 \begin{pmatrix} \langle \Phi_+| \\ \langle \Phi_-| \\ \langle\overline{\Phi}_+| \\ \langle\overline{\Phi}_-| \end{pmatrix} .
 \label{eq:33m}
\ea
\end{widetext}
This Hamiltonian has been obtained using $\hat{H}_0$ and $\hat{H}_c(t_i)$, as given by Eqs.~(\ref{eq:26})
and (\ref{eq:30}), respectively.
Notice that $B_{0+} = (J_z-J_-)$, as it has been proven and explained above.
The eigenvectors (\ref{eq:33m}) can also be computed and read as
%$\bold{v}_1^\top=N ((\alpha+\beta)-\exp(i\xi ),  (\alpha-\beta)+\exp(-i\xi ), 0,  0)$,
%$\bold{v}_2^\top=N ((\alpha-\beta)+\exp(i\xi ), -(\alpha+\beta)+\exp(-i\xi ), 0, 0)$,
%$\bold{v}_3^\top=(0, 0, 1, 0)$,
%$\bold{v}_4^\top=(0, 0, 0, 1)$,
%
\begin{subequations}
\ba
\mathbf{v}_1^\top & = & \frac{1}{N_1} (\alpha-\beta, 1, 0, 0) , \\
\mathbf{v}_2^\top & = & \frac{1}{N_2} (\alpha+\beta, 1, 0, 0) , \\
\mathbf{v}_3^\top & = & (0, 0, 1, 0) , \\
\mathbf{v}_4^\top & = & (0, 0, 0, 1) ,
\ea
\end{subequations}
where $\alpha = -J_-/B_{0+}$ and $\beta=\sqrt{\alpha^2+1}$,
while $N_1 = \sqrt{(\alpha-\beta)^2+1}$   %%%1/(2(\beta^2-\beta\cos(\xi))^{1/2})$.
and $N_2=\sqrt{(\alpha+\beta)^2+1}$ are norm factors.
%where $N$ is the normalization factor, being,
The corresponding eigenvalues are
\begin{subequations}
\ba
 h_z^{(1)} & = & -J_z - B_{0+}\beta , \\
 h_z^{(2)} & = & -J_z + B_{0+}\beta , \\
 h_z^{(3)} & = & J_z - J_+ , \\
 h_z^{(4)} & = & J_z + J_+ ,
\ea
\end{subequations}
%
%%WQ: I think in ew1 and ew2 are interchanged the signs?
%% $h_z^{(1)}=-J_z-B_{0+}\beta$, $h_z^{(2)}=-J_z+B_{0+}\beta$
%
Thus in the quantum Zermelo Hamiltonian, the set of eigenvalues and eigenvectors are time independent as expected.

%%%%%%%%%%%%%%%%%%%%%%%%%%%%%%%%%%%%%%%%%%%%%%%%%%%%%%%%%%%%%%%%%%%%%%%

\subsection{Spin-flip in a Heisenberg dimer}
\label{sec33}

In Sec.~\ref{sec32} we have assumed the functional form of a Zeeman coupling for the
control Hamiltonian, even though the algorithm presented in Sec.~\ref{sec2} does not assume
any particular form for this Hamiltonian.
One may then wonder what would be the resulting control Hamiltonian if its form is not imposed \textit{a priori}.

Let us thus consider that the initial and final states, $|\psi_i\rangle$ and $|\psi_f\rangle$, respectively, are orthonormal.
It is then easy to notice that Eq.~(\ref{eq:22}) reads as
\begin{equation}
 \Delta T = \frac{\pi}{\sqrt{2k}} ,
 \label{eq:22bis}
\end{equation}
i.e., the time needed to reach a target state is inversely proportional to the square root of $k$.
Actually, since $k$ is related to energy, this relation is just a reminiscence of the time-energy
uncertainty relation: the larger the amount of energy put into play to optimally guide the vector
state to its final destination, the shorter the time employed in the journey, and vice versa.
Now, given $\Delta T$, it is then easy to find a general expression for the control Hamiltonian $H_c(t_i)$,
as seen in Sec.~\ref{sec2},
\ba
 \label{control_H_0rtonormal}
  H_c(t_i) & = & i \sqrt{\frac{k}{2}} \left( e^{\pi i\epsilon_{f}/\sqrt{2k}} |\psi_f\rangle \langle \psi_i| \right. \nonumber \\
  & & \qquad \qquad \left. - |\psi_i\rangle \langle \psi_f|e^{-\pi i\epsilon_{f}/\sqrt{2k}} \right),
\ea
where $\epsilon_{f}$ is the energy of the final state $|\psi_f\rangle$.

To gain some insight into the structure of the above control Hamiltonian, we pick up the particular case considered in the previous section, {\it viz.}, the case where initial and final states are maximally entangled Bell states.
Thus, with the choice $|\psi_i\rangle = |\Phi_+\rangle$ and $|\psi_f\rangle = |\Phi_-\rangle$, and hence $\epsilon_{f} = -J_z + J_{-}$, we have
\begin{subequations}
\ba
|\Phi_+\rangle \langle \Phi_-| & = &  \frac{1}{4} \left( \hat{\sigma}_z^{(1)} + \hat{\sigma}_z^{(2)} \right) - \frac{i}{4} \left( \hat{\sigma}_x \otimes \hat{\sigma}_y  + \hat{\sigma}_y \otimes \hat{\sigma}_x \right) , \nonumber \\
 & & \\
|\Phi_-\rangle \langle \Phi_+| & = & \frac{1}{4} \left( \hat{\sigma}_z^{(1)} + \hat{\sigma}_z^{(2)} \right) + \frac{i}{4} \left( \hat{\sigma}_x \otimes \hat{\sigma}_y  + \hat{\sigma}_y \otimes \hat{\sigma}_x \right) . \nonumber \\
 & &
\ea
\end{subequations}
Substituting these expressions into the control Hamiltonian~\eqref{control_H_0rtonormal}, we finally obtain
\ba
 \label{control_H_Bell_k}
 H_c(t_i) & = & \frac{1}{2}\sqrt{\frac{k}{2}} \left[ \sin \left(\epsilon_{f} \frac{\pi}{\sqrt{2k}}\right) (\hat{\sigma}_z^{(1)} + \hat{\sigma}_z^{(2)}) \right. \nonumber \\
 & & \left.  - \cos \left( \epsilon_{f} \frac{\pi}{\sqrt{2k}} \right) (\hat{\sigma}_x\otimes\hat{\sigma}_y + \hat{\sigma}_y\otimes\hat{\sigma}_x ) \right] . \nonumber \\ & &
\ea

From Eq.~\eqref{control_H_Bell_k}, it is clear that the control Hamiltonian adopts the form of a Zeeman coupling for some particular $k$ values, and hence it can be implemented in the laboratory.
More specifically, this is the case when the condition
\begin{equation}\label{k_spin}
 k = \frac{\epsilon_{f}^2}{2\left(n + 1/2 \right)^2} = \frac{(J_z-J_{-})^2}{2\left(n + 1/2 \right)^2},
\end{equation}
is satisfied, with $n \in \mathbb{Z}$.
Accordingly, given $J_z$, the maximum $k$ value is determined from the relation
\begin{equation}
 k = 2\epsilon_{f}^2 = 2(J_z-J_{-})^2,
\end{equation}
which corresponds to the minimum control time,
\begin{equation}
 \Delta T = \frac{\pi}{2|\epsilon_{f}|} = \frac{\pi}{2|J_z-J_{-}|},
\end{equation}
as it follows from \eqref{eq:22bis}.

%%%%%%%%%%%%%%%%%%%%%%%%%%%%%%%%%%%%%%%%%%%%%%%%%%%%%%%%%%%%%%%%%%%%%%%

\subsection{The Cu(II) acetate molecular complex}

As a realistic application of the time-optimal quantum Zermelo navigation, we consider the
copper(II) acetate monohydrate. This complex corresponds to an antiferromagnetic coupled
Heisenberg spin dimer with effective spins $S_1=S_2=1/2$. As such, this system can be cast in
the form of an interacting two-qubit described by a Heisenberg spin dimer as in the previous
section. 
Our goal is to find the optimal time for the transition between two maximally entangled (Bell) states to occur for a physically implementable control Hamiltonian in the form of a Zeeman coupling. 

\begin{figure}[t]
	\centering
	\includegraphics{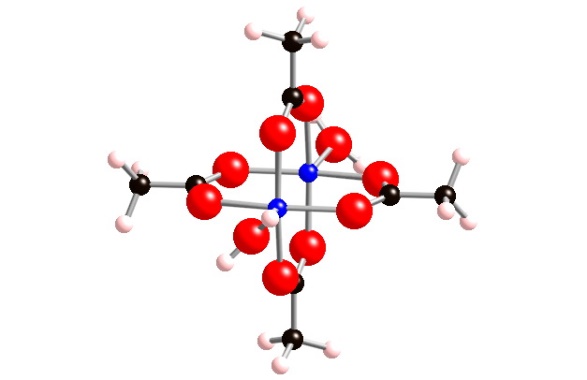}
	\caption{Schematic picture of the paddle-wheel centrosymmetric molecular complex Cu$_2$(O$_2$CCH$_3$)$_4\cdot$2H$_2$O for the crystal structure of copper(II) acetate monohydrate.}
	\label{Fig1}
\end{figure}

The crystal structure of copper(II) acetate monohydrate, Cu$_2$(O$_2$CCH$_3$)$_4\cdot$2H$_2$O,
has been determined by x-ray powder diffraction \cite{niekerk1953} and refined by neutron
diffraction at room temperature \cite{brown1973}.
%, reporting the structural parameters: $a = 13.167(4)$~\AA, $b = 8.563(2)$~\AA, $c = 13.862(7)$~\AA, $\beta = 117.019(2)^\circ$, space group $C2/c4$, $Z=4$.
The crystal is formed by well-defined and separated molecular entities, as displayed in Fig.~\ref{Fig1}.
This complex has a paddle-wheel centrosymmetric structure with two equivalent Cu(II) centers at $2.6143 \dot{\text{A}}$.

Abragam and Bleaney \cite{abragam-bleany-bk} (see p.~503) provide a detailed analysis of the
magnetic interaction in this classical Cu(II) dinuclear antiferromagnetic complex by assuming
the following magnetic Hamiltonian for the isolated dimer of identical spins:
\begin{eqnarray}
 \hat{H} & = & g_z \mu_B H_z \left[ \hat{S}_z^{(1)} + \hat{S}_z^{(2)} \right] + J_x \hat{S}_x^{(1)} \hat{S}_x^{(2)} \nonumber \\
 & & + J_y \hat{S}_y^{(1)} \hat{S}_y^{(2)} + J_z \hat{S}_z^{(1)} \hat{S}_z^{(2)} ,
 \label{eq:23_AB}
\end{eqnarray}
where $g_z$ corresponds to the $z$ component of the $g$ tensor of the magnetic centers having
the same principal axes $(x,y,z)$, and $H_z$ represents the external magnetic field directed
along one of these principal axes (taken here as the $z$ axis).  
It can then be shown that for this molecular complex, the experimental values of the
parameters of the spin Hamiltonian in (\ref{eq:23_AB}) take the
following values: $g_z = 2.43$, $J_x = 297.793$~cm$^{-1}$, $J_y = 297.753$~cm$^{-1}$, and $J_z = 298.453$~cm$^{-1}$.
To cast these values in the Heisenberg spin Hamiltonian for the dimer given in (\ref{eq:23}),
we have to take into account the relation between Pauli matrices $\hat{\sigma}_j^{(i)}$ and
the corresponding spin operators $\hat{S}_j^{(i)}$. In this case, these $J_i$ values have to
be divided by $-$4 and $B^{(i)} = g_z^{(i)} \mu_B H_z^{(i)}/4$ in SI units. Then, with the
choice of $|\psi_i\rangle = |\Phi_+\rangle$ and $|\psi_f\rangle = |\Phi_-\rangle$, and hence
$\epsilon_{f} = -J_z + J_{-}$, the maximum value of $k$ compatible with a Zeeman-type
coupling of the form in (\ref{eq:23_AB}) is $k = 2\epsilon_{f}^2 = 2(J_z-J_{-})^2 \sim 2J_z$,
and the minimum control time corresponds to $\Delta T \sim 0.2$~ps.

%%%%%%%%%%%%%%%%%%%%%%%%%%%%%%%%%%%%%%%%%%%%%%%%%%%%%%%%%%%%%%%%%%%%%%%
%%%%%%%%%%%%%%%%%%%%%%%%%%%%%%%%%%%%%%%%%%%%%%%%%%%%%%%%%%%%%%%%%%%%%%%

\vspace{-0.45cm}

\section{Conclusions}
\label{sec4}

\vspace{-0.15cm}

Given the actual position of a classical particle under the action of a given time-independent force field, there exists an optimal control velocity that, acting constantly on the particle, allows it to reach another position of interest in the least possible time. This problem, known as Zermelo navigation problem~\cite{zermelo31,caratheodory}, can be recast in the realm of quantum mechanics by simply substituting the classical particle by a quantum state. In this context, a time-independent Hamiltonian plays the role of the underlying classical force-field,
and a time-dependent control Hamiltonian with constant energy resource bound is analogous to the control velocity in the classical navigation problem. 
A first solution to the above quantum Zermelo problem was put forth by Russell and Stepney \cite{stepney:PRA:2014,stepney:JPA:2015} and Brody {\it et al.}~\cite{brody15,brogib15} for a particular energy resource bound. Here we have extended this result for general energy resource bounds.

From a fundamental point of view, the solution to the quantum Zermelo problem defines a pair of conjugate variables, {\it viz.}, the energy resource bound and the control time, that minimize the energy-time uncertainty. 
While the time-energy uncertainty relation still arouses controversy, in recent decades there have been several
attempts towards its explanation. This effort has led to
the interpretation of the time-energy uncertainty relation as
a so-called quantum speed limit, i.e., the ultimate
bound imposed by quantum mechanics on the minimal
evolution time between two distinguishable states of a system (see, for instance, the discussion in Ref.~\cite{brogib15}, in direct connection to the quantum Zermelo
navigation problem, or the more recent one \cite{pires2016generalized}, as well as references therein).
Therefore the solution to the quantum Zermelo problem attains the quantum speed limit for any energy resource bound.

In the above respect, however, we have proven that the solution of the quantum navigation problem does not always lead to physically implementable control Hamiltonians. For a single qubit and two interacting qubits, we have shown that energy resources leading to physically implementable control Hamiltonians are not singular but follow a well-defined mathematical pattern. Specifically, for orthogonal initial and target states, the resource energy bound of physically implementable control Hamiltonians does obey a quantization rule that depends, exclusively, on the energy of the target state.

As a realistic application of the time-optimal quantum Zermelo navigation, we have shown results for an acetate molecular complex. The magnetic behavior of copper(II) acetate monohydrate corresponds to an antiferromagnetic ($S1=S2=1/2$) coupled spin dimer. As such, this system can be cast in the form of an interacting two-qubit described by a dimer Heisenberg spin chain. By employing available experimental data we have evaluated the optimal time for the transition between two maximally entangled (Bell) states to occur. For a physically implementable control Hamiltonian in the form of a Zeeman coupling, this time is of the order of a few femtoseconds.   

Finally, we have also shown
that the evolution governed by the Zermelo control Hamiltonian is not adiabatic in general. That is, for an initial state described by a superposition of eigenstates of the full (underlying plus control) Zermelo Hamiltonian, the time evolution governed by the Schr\"odinger equation will not keep constant the population of the system in a given instantaneous eigenstate of the time-dependent Zermelo Hamiltonian.

%%%%%%%%%%%%%%%%%%%%%%%%%%%%%%%%%%%%%%%%%%%%%%%%%%%%%%%%%%%%%%%%%%%%%%%
%%%%%%%%%%%%%%%%%%%%%%%%%%%%%%%%%%%%%%%%%%%%%%%%%%%%%%%%%%%%%%%%%%%%%%%

\vspace{-0.15cm}

\begin{acknowledgments}

\vspace{-0.15cm}

The authors would like to thank Prof. Alberto Castro for fruitful discussions.
Financial support is acknowledged to the Spanish Ministerio de Econom\'{i}a y Competitividad, Project No.\ CTQ2016-76423-P and PID2019-109518GB-I00, Spanish Structures
of Excellence Mar\'{i}a de Maeztu program through grant MDM-2017-0767,
both co-funded by the European Regional Development Fund (ERDF) of the European Union,
and Generalitat de Catalunya, Project No.\ 2017 SGR 348 (J.M.B., G.A., I.P.R.M.);
the Spanish Research Agency (AEI) and the
European Regional Development Fund (ERDF), Grant No.\ FIS2016-76110-P (A.S.);
and the European Union's Horizon 2020 Research and Innovation
Programme under Marie Sklodowska-Curie Grant Agreement BeBOP No.\ 752822 (G.A.).

\end{acknowledgments}
	
%%%%%%%%%%%%%%%%%%%%%%%%%%%%%%%%%%%%%%%%%%%%%%%%%%%%%%%%%%%%%%%%%%%%%%%
%%%%%%%%%%%%%%%%%%%%%%%%%%%%%%%%%%%%%%%%%%%%%%%%%%%%%%%%%%%%%%%%%%%%%%%

%\bibliography{PRR-biblio}

%\end{document}


\end{document}